\documentclass[12pt]{article}
\usepackage{epsfig,amssymb,amsmath,psfrag,multirow,epstopdf,color}
\usepackage{breqn}
\usepackage{placeins}
\usepackage{array}
\usepackage{cite}

\numberwithin{equation}{section}

\newcommand{\be}{\begin{equation}}
\newcommand{\ee}{\end{equation}}
\newcommand{\bea}{\begin{eqnarray}}
\newcommand{\eea}{\end{eqnarray}}

\def\beq{\begin{equation}}
\def\eeq{\end{equation}}

\def\beqn{\begin{eqnarray}}
\def\eeqn{\end{eqnarray}}

\def\bsp#1\esp{\begin{split}#1\end{split}}
\def\spa#1.#2{\left\langle#1\,#2\right\rangle}
\def\spb#1.#2{\left[#1\,#2\right]}

\def\to{\rightarrow}

\newfont{\scyr}{wncyr10 scaled 550}

\def\beq{\begin{equation}}
\def\eeq{\end{equation}}

\begin{document}

%%%%%%%%%%%%%%%%%%%%%%%%%%%%%%%%%%%%%%%%%%%%%%%%%%%%%%%%%%%%%%%%%%%%%%%%
% Shamelessly stolen from Thorsten's thohacks.sty
%%%%%%%%%%%%%%%%%%%%%%%%%%%%%%%%%%%%%%%%%%%%%%%%%%%%%%%%%%%%%%%%%%%%%%%%
\catcode`\@=11
\font\manfnt=manfnt
\def\Watchout{\@ifnextchar [{\W@tchout}{\W@tchout[1]}}
\def\W@tchout[#1]{{\manfnt\@tempcnta#1\relax%
  \@whilenum\@tempcnta>\z@\do{%
    \char"7F\hskip 0.3em\advance\@tempcnta\m@ne}}}
\let\foo\W@tchout
\def\dubious{\@ifnextchar[{\@dubious}{\@dubious[1]}}
\let\enddubious\endlist
\def\@dubious[#1]{%
  \setbox\@tempboxa\hbox{\@W@tchout#1}
  \@tempdima\wd\@tempboxa
  \list{}{\leftmargin\@tempdima}\item[\hbox to 0pt{\hss\@W@tchout#1}]}
\def\@W@tchout#1{\W@tchout[#1]}
\catcode`\@=12
%%%%%%%%%%%%%%%%%%%%%%%%%%%%%%%%%%%%%%%%%%%%%%%%%%%%%%%%%%%%%%%%%%%%%%%%

%%%%%%%%%%%%%%%%%%%%%%%%%%%%%%%%%%%%%%%%%%%%%%%%%%%%

\thispagestyle{empty}

\null\vskip-10pt \hfill
\begin{minipage}[t]{42mm}
 ARPHY-215/16 %\hskip1cm \ \ \ DESY\\
\end{minipage}
\vspace{5mm}

\begingroup\centering
{\Large\bfseries\mathversion{bold}
 On a residual  freedom of  the next-to-leading  BFKL eigenvalue  in color adjoint representation  in   planar $\mathcal{N} = 4$ SYM.
  \par}%
\vspace{7mm}

\begingroup\scshape\large
Sergey~Bondarenko  and Alex~Prygarin  \\
\endgroup
\vspace{5mm}
\begingroup\small
 \emph{ Physics Department, Ariel University, Ariel 40700, Israel} \\
 \endgroup

\vspace{0.4cm}
\begingroup\small
E-mails:\\
{\tt   sergeyb@ariel.ac.il},\ \ \ {\tt alexanderp@ariel.ac.il}
\endgroup
\vspace{0.7cm}

\textbf{Abstract}\vspace{5mm}\par
\begin{minipage}{14.7cm}
We discuss a  residual freedom of the next-to-leading  BFKL eigenvalue that originates  from ambiguity in   redistributing the next-to-leading~(NLO) corrections between the adjoint BFKL eigenvalue and   eigenfunctions in planar $\mathcal{N}=4$ super-Yang-Mills~(SYM) Theory. In terms of the remainder function of the Bern-Dixon-Smirnov~(BDS) amplitude this freedom is translated to reshuffling correction between the eigenvalue and the impact factors in the multi-Regge kinematics~(MRK) in the next-to-leading logarithm approximation~(NLA). We show that the modified NLO BFKL eigenvalue suggested by the authors in Ref.~\cite{BondPrygHermit} can be introduced in the MRK expression for the remainder function by shifting the anomalous dimension in the impact factor in such a way that the  two and three loop remainder function   is left  unchanged to the NLA accuracy.
\end{minipage}\par
\endgroup

\newpage

\newpage

\tableofcontents

%%%%%%%%%%%%%%%%%%%%%%%%%%%%%%%%%%%%%%%%%%%%%%%%%%%%%%%%%%%%%%%%%

\section{Introduction}

The kernel of the BFKL~(Balitsky-Fadin-Kuraev-Lipatov)~\cite{BFKL} equation contains   real and virtual gluon emissions. The virtual gluon emissions are included in the infrared divergent gluon Regge trajectory.
At present, the BFKL kernel is known to the next-to-leading~(NLO)~\cite{ Kuraev:1977fs, Balitsky:1978ic, Fadin:1995dd,Fadin:1995xg,Fadin:1995km,Kotsky:1996xm,Fadin:1996tb,Blumlein:1998ib,DelDuca:2001gu,Kotikov:2000pm,Kotikov:2002ab,Fadin:2007xy,Fadin:2000kx,Fadin:2000hu, Fadin:1998jv, Gerasimov:2010zzb} order for arbitrary color group representation  in both QCD and its supersymmetric extensions. The infrared~(IR) divergences cancel between the real and the virtual part of the kernel than projected on the singlet color state. This cancellation does not happen  for the BFKL kernel in the adjoint representation, but despite being infrared divergent it  can   be useful in some applications, which also determine the way one treats the IR terms.  For example, in the Bartels-Kwiecinski-Praszalowicz~(BKP)~\cite{BKP} approach of interacting reggeized gluons one can remove "halves"  of two gluon trajectories, while in calculations of the corrections to the Bern-Dixon-Smirnov~(BDS)~\cite{BDS} amplitude one removes trajectory of single    reggeized gluon in order to obtain the IR finite expression for the remainder function. The eigenvalue of reduced IR finite BFKL kernel obtained in the "BDS-like" way is commonly known as the adjoint BFKL eigenvalue. The adjoint BFKL eigenvalue at the leading order was calculated by Bartels, Lipatov and Sabio Vera~\cite{sabio2} and its NLO expression was found by Fadin and Lipatov~\cite{LipFadinAdj}. Then its higher loop corrections were calculated  order by order~\cite{Dixon2014iba}  and in all orders from near-collinear limit using integrability techniques as well as a non-trivial analytic continuation from the collinear to Regge kinematics~\cite{BCHS, JamesYorgos}.

 The BFKL approach to the helicity amplitudes in the Regge kinematics was extensively studied over the last years~\cite{sabio1, LP1, LP2,BLP33,BLPCollRegge,Dixon2012yy,Pennington2012zj,CaronHuot2013fea,Hatsuda2014oza,Lipatov2012gk,Bartels:2013dja, Bartels:2013jna, Bartels:2014ppa, Bartels:2014jya, Bartels:2014mka, Fadin:2013sta, Dixon4Loops,  Broedel:2015nfp, Fadin:2015oda, Bargheer:2015djt, Dixon:2015iva, Dixon:2016epj} and was very useful in understanding   higher order corrections of the BFKL eigenvalue and the impact factor. However, already at the next-to-next-to-leading level the adjoint BFKL eigenvalue was shown to have an alerting feature of having a non-vanishing limit as $\nu \to 0$ after setting the conformal spin $n=0$~\cite{Dixon2012yy}, which in not compatible with existence of a constant BFKL eigenfunction. It was shown~\cite{CaronHuot2013fea} that one way to solve this problem is to account for   corrections to the cusp anomalous dimension in the impact factor.
 In the previous paper the authors~\cite{BondPrygHermit} suggested that a more natural way would be  to redefine the notion of the adjoint BFKL eigenvalue exploiting some ambiguity in its definition. This ambiguity is related to the way one removes IR terms as well as how one redistributes the NLO corrections between the eigenvalue and the impact factor.  Moreover, the energy scale in the leading logarithm approximation is not fixed at the leading order and becomes to be important at the NLO level. The authors claimed that there is  enough freedom to modify the adjoint BFKL eigenvalue in such a way that the corresponding BDS remainder function is left intact with the   next-to-leading logarithm accuracy. In the present paper we show in details how the freedom of redistributing NLO corrections between the BFKL eigenvalue and eigenfunctions is realized in the remainder function as interchanging the corresponding corrections between the eigenvalue and the impact factor. The paper is organized as follows. In the next section we discuss the origin of the freedom of redistributing the NLO corrections between the eigenvalue and eigenfunctions of the reduced BFKL kernel. Then we show how this freedom is realized in the BDS remainder function
 modifying the eigenvalue and the impact factors,
 and explain why this procedure does not affect the final expression of the remainder function to the NLA accuracy.

\section{Residual freedom of          BFKL  eigenvalue and  eigenfunctions}\label{functions}
In the multi-Regge kinematics~(MRK) the effective summation parameter is $a \ln \frac{s}{s_0}$, where $s$ is the center of mass energy squared and $s_0$ is some energy scale. The leading contribution is of the order of  $\left(a \ln  s/s_0 \right)^{L-1}$, where $L$ is a loop order in the perturbative expansion, while the next-to-leading contribution is suppressed by one power of log of $s$, namely $a\left(a \ln  s/s_0 \right)^{L-2}$ and commonly is referred to as the next-to-leading logarithm approximation~(NLA). In the present paper we consider only two and three loop BDS  remainder function  $R_6|_{MRK, 2 \to 4}$ to the NLA accuracy  in MRK for the $2\to 4$ gluon scattering. To this accuracy the remainder function is fully determined by the energy scale $s_0$, the leading-order~(LO) and  the next-to-leading order~(NLO)\footnote{For some historical reasons one says that  the scattering amplitude is calculated in the next-to-leading logarithm approximation~(NLA), but the corresponding  BFKL eigenvalue  is the next-to-leading order~(NLO) eigenvalue. Same for impact factors. }  impact factor as well as the LO and the NLO adjoint BFKL eigenvalue in the planar limit. However, there is a residual freedom related to   redistribution of the NLO correction between the BFKL eigenvalue, the impact factor and the energy scale.  This freedom is equivalent to a freedom of  redistributing the  NLO corrections between the BFKL eigenvalue and the eigenfunction in such a way that the BFKL kernel is not affected.
To illustrate this statement we schematically write the reduced infrared finite  BFKL kernel $\mathcal{K}$ in the following form
\beqn
\omega \otimes \phi \otimes \phi^*  =\mathcal{K},
\eeqn
where $\omega$ and $\phi $ are  the  eigenvalue and  the eigenfunction of the BFKL equation.
The reduced infrared finite BFKL kernel is obtained by removing the Regge gluon trajectory from the full infrared divergent BFKL kernel in the adjoint representation of the color gauge group. One uses the reduced adjoint BFKL kernel in calculations of the BDS remainder function   because the gluon Regge trajectory is built in the BDS amplitude by construction.

The eigenfunctions of the LO kernel are also eigenfunctions of the NLO kernel~(cf.\cite{LipFadinNLOsinglet}) for the singlet case.  In the color adjoint BFKL for the reduced kernel the situation is the same, which allowed Fadin and Lipatov  to calculate the NLO eigenvalue~\cite{LipFadinAdj}. We denote it as follows
\beqn
\omega= \omega_{LO} + a \;\omega_{NLO}
\eeqn
and thus according to Ref.~\cite{LipFadinAdj} it reads
\beqn
(\omega_{LO} + a \;\omega_{NLO})\otimes \phi_{LO} \otimes \phi_{LO}^*  =\mathcal{K}_{LO} + a \; \mathcal{K}_{NLO}
\eeqn
for the coupling constant $a=g^2 N_c/ (8 \pi^2)$.
In their previous paper~\cite{BondPrygHermit} the authors argued that the NLO eigenvalue $\omega_{NLO}$ can be modified
 \beqn\label{mod_omega}
 \omega_{NLO}\to \tilde{\omega}_{NLO}=\omega_{NLO}+\Delta \omega_{NLO}
 \eeqn
 to comply with the Hermitian separability properties without affecting the remainder function to this order. This can be done by pushing some of the NLO corrections to the eigenfunction as follows~\footnote{A similar procedure for  NLO eigenfunctions of the singlet BFKL kernel was considered in Refs.~\cite{Bondarenko:2008uc, Chirilli:2013kca}.  }
 \beqn
(\omega_{LO} + a \; \omega_{NLO}+a\;\Delta \omega_{NLO})\otimes (\phi_{LO}+a \; \phi_{NLO}) \otimes (\phi_{LO}+a \; \phi_{NLO})^*  =\mathcal{K}_{LO} + a \; \mathcal{K}_{NLO},
\eeqn
which leaves the kernel $\mathcal{K}_{LO} + a \; \mathcal{K}_{NLO}$ unchanged. The  possibility  of the suggested modification of the NLO eigenvalue  $ \omega_{NLO}\to  \omega_{NLO}+\Delta \omega_{NLO}$ was criticised by Fadin and Fiore~\cite{FadinFioreHermit} based on their calculations using only the LO eigenfunction that is naturally inconsistent with BFKL kernel because
 \beqn
(\omega_{LO} + a \; \omega_{NLO}+a\;\Delta \omega_{NLO})\otimes \phi_{LO}  \otimes  \phi_{LO}^*  \neq \mathcal{K}_{LO} + a \; \mathcal{K}_{NLO}.
\eeqn
The authors    absolutely  agree with the  conclusion made by Fadin and Fiore that the expression
  \beqn
  \;\Delta \omega_{NLO} \otimes  \phi_{LO} \otimes  \phi_{LO}^*
 \eeqn
on its own does not make much sense, but our statement is equivalent to    compensating  this term with the  NLO corrections to the eigenfunction in such  a way that the kernel is left unchanged. In calculus this corresponds to passing to another expansion basis for some function, which does not affect the function itself   rather modifies the coefficients of its expansion, the BFKL eigenvalue in our case. To the best of our knowledge, the uniqueness of the eigenfunction of the NLO BFKL  kernel   was never discussed before. We leave the analysis of the uniqueness and possible  forms of the NLO BFKL eigenfunction for our further publications and only want to show how this residual freedom is realized for the BDS remainder function.

\section{BDS remainder function in multi-Regge kinematics}

The planar BDS amplitude in $\mathcal{N}=4$ SYM possesses correction starting at two loop order of the perturbative expansion.
One of the reasons for this is the fact that the analytic properties of the BDS amplitude are not compatible with  cut singularities   in the complex angular momentum plane, called Mandelstam or Regge cuts. The Mandelstam cut contributions cannot be obtained exponentiating  the one loop result in momentum space, in the way the BDS amplitude is constructed.
For the $2 \to 4$ amplitude with two produced particles $k_1$ and $k_2$  the correction to the BDS amplitude, the so-called remainder function is a function of three  conformal cross ratios $u_i (i=1,2,3)$ in dual momentum space expressed in terms of the Mandelstam invariants as follows
\beqn
u_1=\frac{s s_2}{s_{012} s_{123}}, \;\; u_2=\frac{s_1 t_3}{s_{012} t_2}, \;\; u_3=\frac{s_3 t_1}{s_{123} t_{2}},
\eeqn
where $s$, $s_i$ and $s_{ij}$ are related to the center of mass energies, while $t_i$ stand for the momentum transfer.
In the multi-Regge kinematics~(MRK) $s\gg s_{012}, s_{123} \gg s_1,s_2,s_3 \gg t_1, t_2, t_3 $ the cross ratios greatly simplify
\beqn
1-u_1 \to 0, \; \tilde{u}_2=\frac{u_2}{1-u_1} \propto 1, \; \tilde{u}_3=\frac{u_3}{1-u_1} \propto 1
\eeqn
 and $u_1$ possesses a phase in the Mandelstam region $u_1=|u_1| e^{- i2\pi}$.
The analytic form of the correction to the BDS amplitude in the multi-Regge kinematics  was   introduced by Bartels, Lipatov and Sabio Vera~\cite{sabio2} and its more general form~ \cite{Dixon4Loops, LipFadinAdj}  reads
\beqn\label{R6MRK}
&&\exp\left[R_6+i \pi \delta_{MRK}\right]|_{MRK, 2 \to 4}=\cos \pi \omega_{ab}
\\
&&
\hspace{1cm}
+i \frac{a}{2} \sum_{n=-\infty}^{+\infty}(-1)^n \left(\frac{w}{w^*}\right)^{\frac{n}{2}} \int_{-\infty}^{+\infty} \frac{d \nu }{\nu^2+\frac{n^2}{4}} |w|^{2 i\nu} \; \Phi_{\text{Reg}}(\nu,n) \nonumber
 \; \left(-\frac{1}{1-u_1}\frac{|1+w|^2}{|w|}\right)^{\omega^{adj}(\nu,n)}.
\eeqn
Here $\omega_{ab}$ and $\delta_{MRK}$ are  given by
\beqn
\omega_{ab}=\frac{1}{8} \gamma_K(a) \ln |w|^2, \;\;
\delta_{MRK}= \frac{1}{8} \gamma_K(a) \ln \frac{|w|^2}{|1+w|^4}
\eeqn
and $\gamma_K(a)\simeq 4 a - 4 a^2 \zeta_2+...$  is  the cusp anomalous dimension known to any order in the perturbative expansion~\cite{beisert}.  The complex variable $w$ is related to the  transverse momenta of the produced particles $k_1$, $k_2$ and the momentum transfers $q_1$, $q_2$ and $q_3$   as follows
\beqn
w=\frac{q_3 k_1 }{k_2 q_1}=|w| e^{i \phi_{23}}, \; \; |w|^2=\frac{u_2}{u_3},\;\; \cos \phi_{23}=\frac{1-u_1-u_2-u_3}{2 \sqrt{u_2u_3}}.
\eeqn

The energy dependence in (\ref{R6MRK}) is encoded in (\ref{R6MRK}) by
\beqn
\frac{1}{1-u_1}\frac{|1+w|^2}{|w|}=\frac{s}{s_0}
\eeqn
as well as  the function   $\omega^{adj}(\nu,n)$, which is  the eigenvalue of the reduced infrared finite color adjoint BFKL kernel in the planar $\mathcal{N}=4$ SYM.
The propagation of the BFKL state is then convolved with a product of two impact factors given by
 \beqn
  \frac{ (-1)^n  }{\nu^2+\frac{n^2}{4}}   \; \Phi_{\text{Reg}}(\nu,n).
 \eeqn
To keep the connection with   previous publications we refer to $ \Phi_{\text{Reg}}(\nu,n)$ as the impact factor in the $(\nu,n)$ space.

We are interested only in the NLO corrections and write
 \beqn
 \Phi_{\text{Reg}}(\nu,n)=1+a \;\Phi^{(1)}_{\nu,n}+...
 \eeqn
as well as
 \beqn
 \omega^{adj}=-a \left(E^{(0)}_{\nu,n}+a E^{(1)}_{\nu,n}+...\right),
 \eeqn
 where $E^{(0)}_{\nu,n}$ and  $E^{(1)}_{\nu,n}$ are the LO and NLO  adjoint BFKL eigenvalues and  $\Phi^{(1)}_{\nu,n}$ is the NLO impact factor.

 The  leading order  adjoint BFKL eigenvalue~\cite{sabio2} reads
 \beqn\label{E0}
 E^{(0)}_{\nu,n}=-\frac{1}{2} \frac{|n|}{\nu^2 +\frac{n^2}{4}}+\psi\left(1+i \nu+\frac{|n|}{2}\right)
+\psi\left(1-i \nu+\frac{|n|}{2}\right)-2 \psi(1)
 \eeqn
 and also can be written as follows~(see eq.(85) of Ref.~\cite{sabio2})
 \beqn\label{Ehol0}
 E^{(0)}_{\nu,n}= \frac{1}{2} \left(\psi\left( i\nu +\frac{n}{2}\right)
+
\psi\left(-i\nu +\frac{n}{2}\right)
+
\psi\left( +i\nu -\frac{n}{2}\right)
+
\psi\left( -i\nu -\frac{n}{2}\right)
\right)-2 \psi(1).
\eeqn
 At first sight those two representation of the LO eigenvalue are not the same numerically   and even have different analytic structure, for example setting $\nu=0$ and then taking limit $n \to 0$. However, one should remember that they are always considered under the integral over $\nu$ and the sum over integer $n$, and thus they are equivalent, which can be shown using    the  reflection identity of  the  digamma function.

The adjoint NLO BFKL eigenvalue $E^{(1)}_{\nu,n}$ was calculated by Fadin and Lipatov~\cite{LipFadinAdj}  using LO eigenfunctions from the reduced infrared safe BFKL kernel in the color adjoint representation in planar $\mathcal{N}=4$ SYM.  In the most compact form  it can be written as follows(cf.~\cite{Dixon4Loops})
\beqn\label{E1Dixon}
E^{(1)}_{\nu,n}= -\frac{1}{4}D^2_{\nu} E_{\nu,n} +\frac{1}{2} V D_{\nu} E_{\nu,n}-\zeta_2 E_{\nu,n}-3 \zeta_3
\eeqn
in terms of
\beqn\label{VV}
V \equiv -\frac{1}{2} \left[\frac{1}{i \nu +\frac{|n|}{2}}-\frac{1}{-i \nu +\frac{|n|}{2}}\right]=\frac{i \nu}{\nu^2 +\frac{n^2}{4}}
\eeqn
as well as derivative  defined by $D_\nu=-i \partial_\nu$.

 The NLO impact factor
 \beqn
 \Phi^{(1)}_{\nu,n}=-\frac{1}{2} \left(E^{(0)}_{\nu,n}\right)^2-\frac{3}{8} N^2-\zeta_2,
 \eeqn
 where
 \beqn
N \equiv \text{sgn}(n)\left[\frac{1}{i \nu +\frac{|n|}{2}}+\frac{1}{-i \nu +\frac{|n|}{2}}\right] =\frac{n}{\nu^2+\frac{n^2}{4}},
\eeqn
 was calculated by Lipatov and one of the authors~\cite{LP2}  analytically continuing  the exact two loop remainder function found by Goncharov, Spradlin, Volovich and Vergu~\cite{GSVV} to the Mandelstam region, and then was redefined by Fadin and Lipatov~\cite{LipFadinAdj} to fit the energy scale in agreement with the Regge factorization property.
In the next section we review   properties of the adjoint  NLO BFKL eigenvalue and discuss  a residual freedom of redistributing NLO corrections between the eigenvalue and the corresponding impact factor.

\section{Hermitian separability and  transition from singlet to adjoint BFKL eigenvalue }

 In Ref.~\cite{BondPrygHermit} the authors argued that it is possible to   modify the NLO eigenvalue
 \beqn\label{Emod}
 E^{(1)}_{\nu,n} \to  E^{(1)}_{\nu,n}+ \Delta   E^{(1)}_{\nu,n}
\eeqn
 accompanied by a change of   the impact factor and the energy scale in such  way that the remainder function in (\ref{R6MRK}) remains intact to the next-to-leading logarithm~(NLA) accuracy in MRK.
 This modification of the NLO eigenvalue of the BFKL kernel in the color adjoint representation was needed to have the Hermitian separability and to establish a non-trivial connection with the  corresponding NLO eigenvalue in the color singlet state. The authors also suggested the exact form of  $\Delta   E^{(1)}_{\nu,n}$, namely

 \beqn\label{DeltaEbare}
 &&\Delta   E^{(1)}_{\nu,n} =
 \frac{1}{2}\left[\psi\left(1+ i \nu +\frac{|n|}{2}\right)-\psi\left(1- i \nu +\frac{|n|}{2}\right)\right]
\\
&&
\times
\left[
-\frac{i \nu  \left| n\right| }{\left(\nu ^2+\frac{n^2}{4}\right)^2}
+
\psi'\left(1+ i \nu +\frac{|n|}{2}\right)-\psi'\left(1- i \nu +\frac{|n|}{2}\right)\right]. \nonumber
\eeqn

Below we review the main results of Ref.~\cite{BondPrygHermit} and discuss the motivation for the modification of the adjoint NLO eigenvalue given by  (\ref{Emod}) and (\ref{DeltaEbare}). The BFKL equation describes a bound state of two reggeized gluons in an arbitrary color state. The BFKL equation is a Schr\"{o}dinger type equation, with an eigenvalue being a function of the anomalous dimension $\nu$~\footnote{The anomalous dimension of the twist-2 operators is usually denoted  by $\gamma=1/2 +i\nu$ in the BFKL approach. In our analysis, we deal with the integration variable $\nu$, which we call  for simplicity "the anomalous dimension".} and   the conformal spin $n$. By the BFKL eigenvalue one typically means the eigenvalue of the BFKL equation projected on colorless state, called singlet state. It is well known that if BFKL is projected on color adjoint state (color state of one gluon) it reduces to one reggeized gluon if both reggeized gluons in the bound state attached to the same object, e.g. quark line. This does not happen in the six-particle helicity amplitudes, where the two reggeized gluons in the bound state are attached to different vertices resulting in color adjoint BFKL equation~(for more details see Ref.~\cite{sabio2}).

Both the singlet and the adjoint   leading order  BFKL equations are solved exploiting conformal invariance, in the coordinate space for the singlet BFKL and in the dual momentum space for the adjoint BFKL. The conformal groups are defined in quite different spaces, but the LO eigenfunctions and eigenvalues are very similar. Roughly speaking the LO eigenfunctions of the singlet BFKL equation are $\rho^{i\nu+n/2}\bar{\rho}^{i\nu+n/2}$, while for the adjoint BFKL the eigenfunctions  are  $k^{i\nu+n/2}\bar{k}^{i\nu+n/2}$, where $\rho$ and $k$ are the complex coordinate and momentum correspondingly.~\footnote{The actual form of the eigenfunctions is a bit more involved, it accounts for momentum transfer and includes a proper normalization.}

The singlet BFKL eigenvalue $\chi(\nu,n)$ and  adjoint BFKL eigenvalue $E_{\nu,n}$  are also very similar in the leading order. They are both   built of digamma function of an argument shifted by $1/2$.
 Namely, the singlet BFKL leading order reads
\small
\beqn\label{chi01}
&& \chi(n, \gamma) =-\frac{1}{2} \left(\psi\left(\frac{1}{2}+i\nu +\frac{n}{2}\right)
+
\psi\left(\frac{1}{2}-i\nu +\frac{n}{2}\right)
+
\psi\left(\frac{1}{2}+i\nu -\frac{n}{2}\right)
+
\psi\left(\frac{1}{2}-i\nu -\frac{n}{2}\right)  \nonumber
\right)+2 \psi(1)
\eeqn
\normalsize
while the adjoint BFKL eigenvalue  is given by
\beqn\label{Ehol0}
&&E^{(0)}_{\nu,n}  =\frac{1}{2} \left(\psi\left( i\nu +\frac{n}{2}\right)
+
\psi\left(-i\nu +\frac{n}{2}\right)
+
\psi\left( +i\nu -\frac{n}{2}\right)
+
\psi\left( -i\nu -\frac{n}{2}\right)  \nonumber
\right)-2 \psi(1).
\eeqn

The next-to-leading eigenvalues calculated using LO eigenfunctions are quite different.  The adjoint NLO eigenvalue is constructed solely of polygamma functions and their derivatives, while the singlet NLO  eigenvalue is built of a new type of function, Lerch transcendent and its generalizations.
Another difference between the two is that the NLO singlet eigenvalue can be written in the form of the Bethe-Salpeter equation separating holomorphic $i\nu +n/2$ and antiholomorphic $i\nu-n/2$ coordinates as was shown by Kotikov and Lipatov~\cite{Kotikov:2002ab}, while the adjoint eigenvalue in the original form calculated by Fadin and Lipatov~\cite{LipFadinAdj} does not possess this property called Hermitian separability as it was shown in Ref.~\cite{BondPrygHermit}.  To make the adjoint NLO eigenvalue to comply with the Hermitian separability property the authors suggested in Ref.~\cite{BondPrygHermit} to modify the adjoint NLO eigenvalue adding a term given by (\ref{DeltaEbare}). In the present paper we show that this modification can be introduced without affecting the resulting remainder function to the  NLO accuracy.

The authors also argued in Ref.~\cite{BondPrygHermit} that their proposal of modifying the adjoint NLO is supported by  an observation that the modified adjoint NLO
BFKL eigenvalue can be reproduced by \emph{ad hoc} procedure of replacing sign alternating sums in the singlet NLO expression by sums of constant sign accompanied by a shift of  $1/2$ in the  argument.  This transition from singlet to the adjoint BFKL eigenvalue is of yet unknown nature and is still to be checked for validity at higher orders.   However,  many higher order corrections to the adjoint eigenvalue available  as well as  the  recently calculated NNLO singlet eigenvalue for $n=0$ show the same structure, namely singlet eigenvalue is built of sign alternating sums with only one negative index, while the adjoint eigenvalue is  constructed   of polygamma functions and their derivatives.  This hints a possibility that this rather simple 
prescription for the transition from the singlet to the adjoint eigenvalue is valid at higher orders as well.

In the next section we show how the suggested modification of the adjoint NLO eigenvalue can be introduced without changing the remainder functions to the required NLA accuracy.

\section{Rescaling anomalous dimension}
 Below we show that the required modification of the NLO eigenvalue can be made by changing the anomalous dimension as follows
\beqn\label{nushift}
\nu \to \nu +a\; f^{(1)}_{\nu,n}
\eeqn
where
\beqn\label{f1}
 f^{(1)}_{\nu, n}=\frac{1}{2 i} \left[\psi\left(1+ i \nu +\frac{|n|}{2}\right)-\psi\left(1- i \nu +\frac{|n|}{2}\right)\right]
 =\text{Im}\left[\psi\left(1+ i \nu +\frac{|n|}{2}\right)\right].
 \eeqn
The change of the integration variable $\nu$ does not change the integral in (\ref{R6MRK}) thus satisfying the  condition that one   modifies the NLO eigenvalue $E^{(1)}_{\nu,n}$ leaving the remainder function $R_{6}$ intact to the NLA order in MRK.

%%%%%%%%%%%%%%%%%%%%%%%%%%%%%%%%%%%%%%%%%%
\begin{figure}[t!]
\begin{center}
\includegraphics[width=5.0in]{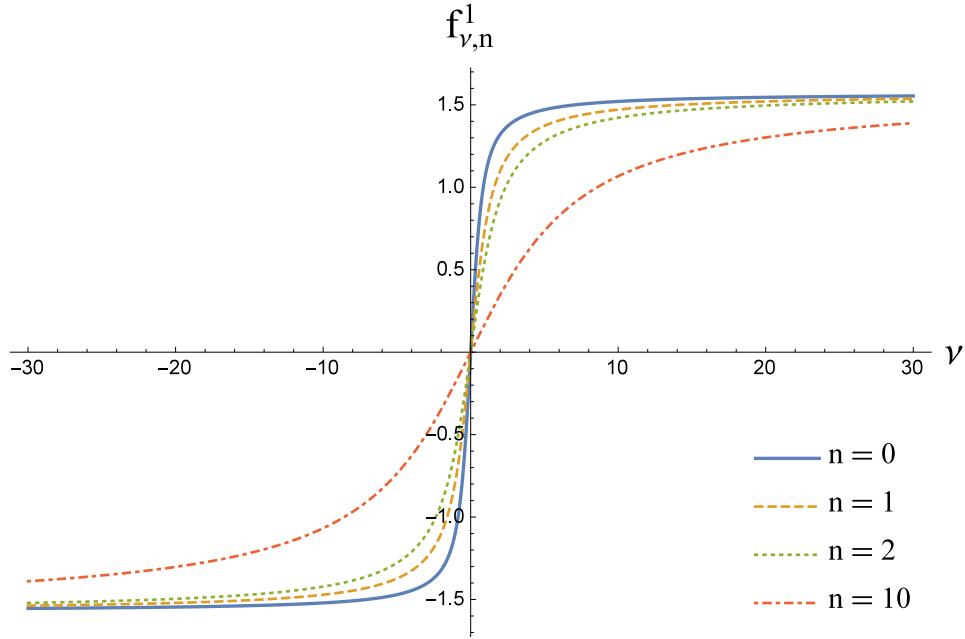}
\end{center}
\caption{The plot of    $f^{(1)}_{\nu,n}$  as a  function of $\nu$ for different values of the conformal spin $n$.  }
\label{fig:f1}
\end{figure}

One can see from Fig.~\ref{fig:f1} that the function $f^{(1)}_{\nu,n}$ is well behaved in the region of the integration over $\nu$ and is limited by $\pm \pi/2$,
which makes it to be  compatible with the perturbative expansion in $a$ and   negligible  for large values of $\nu$ in change of variable $\nu \to \nu + a f^{(1)}_{\nu,n}$.
In particular, for $n=0$  its asymptotic behaviour is given by
\beqn
  f^{(1)}_{\nu,n}|_{n=0} \underset{\nu \to \pm \infty}{\simeq}  \pm \frac{\pi}{2}-\frac{1}{2 \nu}+\mathcal{O}\left(\frac{1}{\nu^2}\right).
\eeqn

Other useful features of $f^{(1)}_{\nu,n}$ are that it is real and antisymmetric in $\nu$ as well as the fact that
\beqn\label{f1exp}
\nu+a f^{(1)}_{\nu,n}|_{n=0} \underset{\nu \to 0}{\simeq} \nu (1+ a \;\zeta_2)+\mathcal{O}(\nu^2).
\eeqn
The expansion in (\ref{f1exp}) seems to compensate  the NLO correction to the cusp anomalous dimension
\beqn
\frac{\gamma_K}{4 a }(\nu+a f^{(1)}_{\nu,n}|_{n=0})  \underset{\nu \to 0}{\simeq} \nu+ \mathcal{O}(\nu^2),
\eeqn
and is likely    related to a way one  removes the infrared divergent    Regge gluon trajectory from the color adjoint  BFKL kernel.

In the Appendix we demonstrate  how the residual freedom of redistributing NLO corrections is realized for the remainder function in multi-Regge kinematics.
We show that a simple change of the integration variable given by  (\ref{nushift}) can be used to redistribute the NLO corrections between that eigenvalue and the impact factor in such a way that  the resulting expression naturally incorporates the modified NLO BFKL eigenvalue in (\ref{DeltaEbare}) while   leaving the whole expression for the remainder function  at two and three loops intact  to the required NLA accuracy.

As it is shown in the Appendix, introducing the change of variable  given by (\ref{nushift}) we can  write the expression for the remainder function in (\ref{R6MRK}) in the following form
\beqn\label{R6Mod2}
&&\exp\left[R_6+i \pi \delta_{MRK}\right]|_{MRK, 2 \to 4}=\cos \pi \omega_{ab}
\\
&&
\hspace{0.2cm}
+i \frac{a}{2} \sum_{n=-\infty}^{+\infty}(-1)^n \left(\frac{w}{w^*}\right)^{\frac{n}{2}} \int_{-\infty}^{+\infty}d \nu \frac{|w|^{2 i\left(\nu+a f^{(1)}_{\nu,n}\right)} }{\left(\nu+a f^{(1)}_{\nu,n}\right)^2+\frac{n^2}{4}}  \; \tilde{\Phi}_{\text{Reg}}(\nu,n) \nonumber
 \; \left(-\frac{1}{1-u_1}\frac{|1+w|^2}{|w|}\right)^{\tilde{\omega}^{adj}(\nu,n)}.
\eeqn
 The  modified NLO BFKL eigenvalue reads
  \beqn\label{omega_tilde}
 \tilde{\omega}^{adj}=-a \left(E^{(0)}_{\nu,n}+a  E^{(1)}_{\nu,n}+a \Delta E^{(1)}_{\nu,n}+...\right),
 \eeqn
where   $ \Delta E^{(1)}_{\nu,n}$ is given by (\ref{DeltaEbare}), and  the modified  NLO impact factor is expressed through
 \beqn
    \tilde{\Phi}_{\text{Reg}}(\nu,n)=1+a \;\Phi^{(1)}_{\nu,n}+a \Delta\;\Phi^{(1)}_{\nu,n} ...
  \eeqn
where $ \Delta \;\Phi^{(1)}_{\nu,n}$  is defined by
 \beqn\label{DeltaPhi}
 \Delta \;\Phi^{(1)}_{\nu,n}=  \partial_\nu f^{(1)}_{\nu,n}=\frac{1}{2}\left[\psi'\left(1+ i \nu +\frac{|n|}{2}\right)+\psi'\left(1- i \nu +\frac{|n|}{2}\right)\right]=\text{Re} \left[\psi'\left(1+ i \nu +\frac{|n|}{2}\right)\right].\;
  \eeqn

The new integral representation  of the MRK remainder function in (\ref{R6Mod2}), which naturally incorporates the modified NLO BFKL eigenvalue presents the main result of   this manuscript. It is worth emphasizing that the remainder function at two and three loops remains the same with the NLA accuracy and not affected by redistribution of the NLO corrections between the eigenvalue and the impact factor.

It is clear that the modification  of the eigenvalue given in (\ref{omega_tilde}) does not change the whole expression for the remainder function to the required NLA accuracy, because it merely follows  from the change of the integration variable. On the other hand the meaning of the variable $\nu$ has been changed and it is related to the original $\nu$ through $\nu \to \nu+a f^{(1)}_{\nu,n}$. Those two coincide at the leading order and slightly differ at the next-to-leading order. The function $f^{(1)}_{\nu,n}$ is limited by $\pm \frac{\pi}{2}$ for any value of the original  $\nu $ and $n$ and thus is significant only in a small region of the  integration over $\nu$ even for reasonably large values of the coupling constant in the perturbative expansion. The original meaning of $n$ as a conformal spin  remains the same and is not affected by the change of $\nu$. The dependence of the anomalous dimension on the conformal spin through $f^{(1)}_{\nu,n}$ can be interpreted as a breaking  of the axial symmetry in the complex  $(w, w^*)$ plane by causing  the dilation to  depend on the angle.

\section{Conclusions and outlook }

In this paper we discuss a residual freedom of redistributing next-to-leading order corrections between the eigenvalue and impact factors for color adjoint BFKL in planar $\mathcal{N}=4$ SYM. This freedom originates from an arbitrariness in solving the BFKL equation by either the eigenfunctions of the LO BFKL kernel or some other eigenfunctions, that can possess   NLO corrections.
The full solution is then expanded in the basis of the new  eigenfunctions and therefore the eigenvalue, being a coefficient of this expansion,
is modified under a change of the expansion basis.

We showed that this residual freedom can be exploited to rewrite the two and three loop expression for the remainder function at the  NLA accuracy in MRK in such a way that it naturally incorporates the modified NLO BFKL eigenvalue suggested by the authors in Ref.~\cite{BondPrygHermit}  and is given by (\ref{Emod}) together with (\ref{E1Dixon})  and (\ref{DeltaEbare}). This modification requires some  redefinition of the NLO impact factor in the $(\nu,n)$ space. The impact factor in the $(\nu,n)$ space is a convolution of the impact factor in the momentum space and the BFKL eigenfunction and it is natural to expect that a change of the eigenfunction is translated into a change of this convolution even  for same the impact factor in the momentum space.  The modifications to the eigenvalue and the impact factor are of the same origin and therefore they  cancel each other leaving the expression for the remainder function unchanged.

The main motivation for modifying the NLO BFKL eigenvalue in the color adjoint state was to restore the Hermitian separability property present in the singlet case. It also helped to establish a non-trivial connection between the adjoint and the singlet  NLO eigenvalues   effectively replacing the alternating sums  in singlet eigenvalue by sums of the constant sign.
This connects between the cylindrical topology of the singlet BFKL and  the plane topology of the adjoint BFKL.

The empirical recipe of  the transition from the singlet to the adjoint eigenvalue may serve a powerful tool for higher loop calculations provided it holds beyond NLO order. It is very encouraging that the known corrections to the adjoint BFKL eigenvalue can be expressed in terms of the constant sign sums while the only known  singlet NNLO BFKL eigenvalue (for zero value of the conformal spin) calculated by   N.~Gromov, F.~Levkovich-Maslyuk and G.~Sizov~\cite{GromovNNLO}\footnote{See also a parallel calculation by Velizhanin~\cite{velizh}.}
 is written in terms of  alternating  harmonic sums with only one negative index.

The residual freedom of redistributing higher order corrections is also present beyond NLA accuracy and is not fixed by the choice of the NLO eigenvalue and the impact factors presented in this paper. This freedom can be exploited to establish  a connection between the  eigenvalues of the BFKL kernel  in color  singlet and adjoint states to all orders in the perturbative expansion.

\section{Acknowledgements}
 The authors are thankful to Jochen Bartels, Victor Fadin and Lev Lipatov  for enlightening discussions on the BFKL Physics.  A.P. is especially indebted to Jochen Bartels for his hospitality at Hamburg, where this project was initiated.

\newpage

\appendix
%\addcontentsline{toc}{section}{aaa}
\section{ Modifying next-to-leading eigenvalue and impact factor }
%\large{\textbf{\hspace{0.4cm} Modifying next-to-leading eigenvalue and impact factor }} \newline \normalsize

In this section we  demonstrate  how the residual freedom of redistributing NLO corrections is realized for the remainder function in multi-Regge kinematics.
In the following we show that a simple change of the integration variable given by  (\ref{nushift}) can be used to redistribute the NLO corrections between that eigenvalue and the impact factor in such a way that  the resulting expression naturally incorporates the modified NLO BFKL eigenvalue in (\ref{DeltaEbare}) while   leaving the whole expression for the remainder function  at two and three loops intact  to the required NLA accuracy.

We write the relevant integral in  (\ref{R6MRK}) as follows
\beqn\label{Aintnu}
&& \int_{-\infty}^{+\infty} \frac{d \nu }{\nu^2+\frac{n^2}{4}} |w|^{2 i\nu} \; \Phi_{\text{Reg}}(\nu,n)
 \; \left(-\frac{1}{1-u_1}\frac{|1+w|^2}{|w|}\right)^{\omega^{adj}(\nu,n)}
 \\
 &&=
 \int_{-\infty}^{+\infty} \frac{d \nu }{\nu^2+\frac{n^2}{4}} |w|^{2 i\nu} \; \Phi_{\text{Reg}}(\nu,n)
 \exp\left[- \omega^{adj}(\nu,n)\left(\ln (1-u_1)+i\pi+F(w)\right)\right],\nonumber
  \eeqn
where
\beqn
F_w=\frac{1}{2}\ln \frac{|w|^2}{|1+w|^4}.
\eeqn

Expanding the integrand of (\ref{Aintnu}) to the NLA order we  get
 \begin{eqnarray}\label{Aexpexp}
&& a \;\Phi_{Reg}(\nu,n) \exp\left[{-\omega^{a} (\ln(1-u_1)+F_{w}+i\pi)}\right] \nonumber
 \simeq a+ a^2 \ln(1-u_1)E^{(0)}_{\nu,n} \\
 &&+ a^2 \left( E^{(0)}_{\nu,n} F_{w}+\Phi^{(1)}_{\nu,n}+i\pi E^{(0)}_{\nu,n} \right) +\frac{a^3}{2}\left(\ln(1-u_1) E^{(0)}_{\nu,n}\right)^2
 \\
 &&+a^3  \ln(1-u_1)\left( E^{(1)}_{\nu,n}+i\pi  \left(E^{(0)}_{\nu,n}\right)^2  + E^{(0)}_{\nu,n} \left( E^{(0)}_{\nu,n} F_{w}+\Phi^{(1)}_{\nu,n}\right)\right), \nonumber
 \end{eqnarray}
where $F_{w}=\frac{1}{2} \ln \frac{|w|^2}{|1+w|^4}$ is a function of transverse momentum, while $E^{(0)}_{\nu,n} $, $E^{(1)}_{\nu,n}$ and $\Phi^{(1)}_{\nu,n}$ are functions of  $\nu$ and $n$.

Next we substitute the integration variable in (\ref{Aintnu})  as follows
\beqn\label{Anushift}
\nu \to \nu +a f^{(1)}_{\nu,n}.
\eeqn

It is convenient to consider the effect of this substitution term  by term. First we note that
\beqn
\left.E^{(0)}_{\nu,n} \right\vert_{\nu \to \nu+ a f^{(1)}_{\nu,n}} \simeq  E^{(0)}_{\nu,n} +
i a f^{(1)}_{\nu,n} D_\nu E^{(0)}_{\nu,n},
\eeqn
where we define  $ D_\nu =- i\partial_\nu$.
 Plugging this into (\ref{Aexpexp}) we get

 \begin{eqnarray}
&& a \;\Phi_{Reg}(\nu,n) \exp\left[{-\omega^{a} (\ln(1-u_1)+F_{w}+i\pi)}\right]|_{\nu \to \nu +a f^{(1)}_{\nu,n}}\nonumber
 \simeq a+ a^2 \ln(1-u_1)E^{(0)}_{\nu,n} \\
 &&+ a^2 \left( E^{(0)}_{\nu,n} F_{w}+\Phi^{(1)}_{\nu,n}+i\pi E^{(0)}_{\nu,n} \right) +\frac{a^3}{2}\left(\ln(1-u_1) E^{(0)}_{\nu,n}\right)^2
 \\
 &&+a^3  \ln(1-u_1)\left( E^{(1)}_{\nu,n}+\Delta   E^{(1)}_{\nu,n} +i\pi  \left(E^{(0)}_{\nu,n}\right)^2  + E^{(0)}_{\nu,n} \left( E^{(0)}_{\nu,n} F_{w}+\Phi^{(1)}_{\nu,n}\right) \right), \nonumber
 \end{eqnarray}
 where
\beqn\label{ADeltaE}
 &&\Delta   E^{(1)}_{\nu,n} =  i f^{(1)}_{\nu,n} D_\nu E^{(0)}_{\nu,n}=\frac{1}{2}\left[\psi\left(1+ i \nu +\frac{|n|}{2}\right)-\psi\left(1- i \nu +\frac{|n|}{2}\right)\right]D_\nu E^{(0)}_{\nu,n} \nonumber
\\
&&
=\frac{1}{2}\left[\psi\left(1+ i \nu +\frac{|n|}{2}\right)-\psi\left(1- i \nu +\frac{|n|}{2}\right)\right]
\\
&&
\times
\left[
-\frac{i \nu  \left| n\right| }{\left(\nu ^2+\frac{n^2}{4}\right)^2}
+
\psi'\left(1+ i \nu +\frac{|n|}{2}\right)-\psi'\left(1- i \nu +\frac{|n|}{2}\right)\right], \nonumber
\eeqn
which gives exactly the modification of the NLO eigenvalue suggested by the authors~(cf.~\cite{BondPrygHermit}), namely
\beqn
 E^{(1)}_{\nu,n} \to  E^{(1)}_{\nu,n}+ \Delta   E^{(1)}_{\nu,n}.
\eeqn
Other terms in the integrand of (\ref{Aintnu}) are also affected by the substitution (\ref{Anushift}). In particular,
we write
\beqn\label{Adrob}
\left.\frac{|w|^{i 2 \nu } }{\nu^2+\frac{n^2}{4}} \right\vert_{\nu \to \nu +a f^{(1)}_{\nu,n}}=\frac{|w|^{i 2 \left(\nu+a f^{(1)}_{\nu,n}\right) } }{\left(\nu+a f^{(1)}_{\nu,n}\right)^2+\frac{n^2}{4}}.
\eeqn
For   clarity of presentation we choose not to expand the expression in (\ref{Adrob}).

The  Jacobian gives
\beqn\label{Ajacob}
d \nu \to   \left(1+ a   \partial_\nu f^{(1)}_{\nu,n} \right) d\nu
\eeqn
and its easy to see that we can absorb the second term in the brackets of (\ref{Ajacob}) in the redefinition of
 the NLO impact factor\footnote{In fact, what we call the NLO impact factor in the $(\nu,n)$ space is a product of $\Phi^{(1)}_{\nu,n}$ and the   expression in (\ref{Adrob}) times $(-1)^n$. } as follows
 \beqn
    \tilde{\Phi}_{\text{Reg}}(\nu,n)=1+a \;\Phi^{(1)}_{\nu,n}+a \Delta\;\Phi^{(1)}_{\nu,n} ...
  \eeqn
where
  \beqn\label{ADeltaPhi}
 \hspace{-0.5cm}\Delta \;\Phi^{(1)}_{\nu,n}=  \partial_\nu f^{(1)}_{\nu,n}=\frac{1}{2}\left[\psi'\left(1+ i \nu +\frac{|n|}{2}\right)+\psi'\left(1- i \nu +\frac{|n|}{2}\right)\right]=\text{Re} \left[\psi'\left(1+ i \nu +\frac{|n|}{2}\right)\right].\;
  \eeqn

Finally, combining  the expanded terms we write the expression for the remainder function
\beqn\label{AR6Mod2}
&&\exp\left[R_6+i \pi \delta_{MRK}\right]|_{MRK, 2 \to 4}=\cos \pi \omega_{ab}
\\
&&
\hspace{0.2cm}
+i \frac{a}{2} \sum_{n=-\infty}^{+\infty}(-1)^n \left(\frac{w}{w^*}\right)^{\frac{n}{2}} \int_{-\infty}^{+\infty}d \nu \frac{|w|^{2 i\left(\nu+a f^{(1)}_{\nu,n}\right)} }{\left(\nu+a f^{(1)}_{\nu,n}\right)^2+\frac{n^2}{4}}  \; \tilde{\Phi}_{\text{Reg}}(\nu,n) \nonumber
 \; \left(-\frac{1}{1-u_1}\frac{|1+w|^2}{|w|}\right)^{\tilde{\omega}^{adj}(\nu,n)}
\eeqn
 for modified NLO BFKL eigenvalue
  \beqn\label{Aomega_tilde}
 \tilde{\omega}^{adj}=-a \left(E^{(0)}_{\nu,n}+a  E^{(1)}_{\nu,n}+a \Delta E^{(1)}_{\nu,n}+...\right),
 \eeqn
where   $ \Delta E^{(1)}_{\nu,n}$ is given by (\ref{ADeltaE}),
and for the modified  NLO impact factor
 \beqn
    \tilde{\Phi}_{\text{Reg}}(\nu,n)=1+a \;\Phi^{(1)}_{\nu,n}+a \Delta\;\Phi^{(1)}_{\nu,n} ...
  \eeqn
where $ \Delta \;\Phi^{(1)}_{\nu,n}$  is defined in (\ref{ADeltaPhi}).

Strictly speaking, expanding the integral in (\ref{Aintnu})  in powers of the coupling constant $a$ after the substitution $\nu \to \nu + a f^{(1)}_{\nu,n}$ we have  also to expand   the upper and lower limits of the integral, which are now some  functions of the coupling constant $a$. However, as it was already mentioned, both  the function $f^{(1)}_{\nu,n}$ and its derivative are limited as $\nu \to \pm \infty$ and thus the terms coming from the expansion of the upper and lower limits vanishes for any finite value of $a$. Another fine point is the case of $n=0$, which has to be considered with a special care because of the infrared divergence  and should be  understood as the principal value of the integral over $\nu$  as it was shown  for one loop in the Appendix of Ref.~\cite{sabio2}.

The new integral representation  of the MRK remainder function in (\ref{AR6Mod2}), which naturally incorporates the modified NLO BFKL eigenvalue presents the main result of   this manuscript. It is worth emphasizing that the remainder function at two and three loops remains the same with the NLA accuracy and not affected by redistribution of the NLO corrections between the eigenvalue and the impact factor.

 One can consider other ways of writing (\ref{AR6Mod2}), for example expanding some terms in (\ref{Adrob}). This would redefine the energy scale  function $F_w$ and further change the impact factor, but bring no new insight into the problem under discussion making the result less transparent.

\newpage
%%%%%%%%%%%%%%%%%%%%%%%%%%%%%%%%%%%%%%%%%%%%%%%%%%%%%%%%%%%%%%%%%%%%%%%%

\end{document}